\documentstyle[epsf]{llncs}
\newfont{\titelfont}{cmr10 scaled 1728}
\newfont{\titelbffont}{cmbx10 scaled 2074}
\newfont{\titelbigfont}{cmr10 scaled 2488}
\def\typeset{}
\begin{document}
\title{Simulating the Effect of Decoherence and Inaccuracies on a Quantum Computer}
\author{Kevin M. Obenland and Alvin M. Despain [obenland,despain]@isi.edu}
\institute{Information Sciences Institute}
\maketitle

\begin{abstract}
A Quantum Computer is a new type of computer which can solve problems such as factoring and 
database search very efficiently. The usefulness of a quantum computer is limited by the 
effect of two different types of errors, decoherence and inaccuracies. 
In this paper we show the results of simulations of a quantum computer which consider both 
decoherence and inaccuracies. We simulate circuits which factor the numbers 15, 21, 35, and 
57 as well as circuits which use database search to solve the circuit satisfaction problem. 
Our simulations show that the error rate per gate is on the order of \(10^{-6}\) for 
a trapped ion quantum computer whose noise is kept below \(\pi/4096\) per gate and with 
a decoherence rate of \(10^{-6}\). 
This is an important bound because previous studies have shown that a quantum computer can 
factor more efficiently than a classical computer if the error rate is of order 
\(10^{-6}\).
\end{abstract}

\subsubsection*{Keywords:}
Quantum Simulation, Ion Trap, Factoring, Database Search

\section{Introduction}

A quantum computer consists of atomic particles which obey the laws of quantum mechanics
\cite{tu:ho}\cite{lloy}. The complexity of a quantum system is exponential with respect 
to the number of particles. Performing computation using these quantum particles results 
in an exponential amount of calculation in a polynomial amount of space and time 
\cite{feyn}\cite{deut}. This quantum parallelism is only applicable in a limited domain. 
Prime factorization is one such problem which can make effective use of quantum 
parallelism\cite{shor}. This is an important problem because the security of the RSA public-key 
cryptosystem relies on the fact that prime factorization is computationally difficult\cite{risa}.

Errors limit the effectiveness of any physical realization of a quantum computer. A quantum 
computer is subject to two different types of errors, decoherence and inaccuracies. 
Decoherence occurs when a quantum computer interacts with the environment. This interaction 
destroys the quantum parallelism by turning a quantum calculation into a classical one. 
The other type of error, inaccuracies in the implementation of gate operations, accumulates 
over time and destroys the results of the calculation.

In this paper we show results of simulations of a quantum computer which is subject to both 
decoherence and inaccuracies. These simulations assume the trapped ion model of a quantum 
computer proposed by Cirac and Zoller\cite{ci:zo}. We study Shor's factorization algorithm by 
simulating circuits which factor the numbers 15, 21, 35, and 57\cite{shor}. We also simulate 
Grover's database search algorithm with a circuit which solves the circuit satisfaction 
problem\cite{grov}. 
The rest of this section gives a brief overview of quantum computers.


%
\subsection{Qubits and Quantum Superposition}
The basic unit of storage in a Quantum Computer is the {\em qubit}. A qubit is like a classical bit 
in that it can be in two states, zero or one. The qubit differs from the classical bit in 
that, because of the properties of quantum mechanics, it can be in both these states 
simultaneously\cite{fe:ls}. A qubit which contains both the zero and one values is said to be 
in the superposition of the zero and one states. The superposition state persists until we 
perform an external measurement. This measurement operation forces the state to one of the 
two values. Because the measurement determines without doubt the value of the qubit, we must 
describe the possible states which exist before the measurement in terms of their probability 
of occurrence. These qubit probabilities must always sum to one because they represent all 
possible values for the qubit.

The quantum simulator represents the qubits of the computer using a complex vector space. Each state in the 
vector represents one of the possible values for the qubits. The bit values of a state are encoded as the index of that state in the vector. The simulator represents each encoded 
bit string with a non zero amplitude in the state vector. The probability of each state is 
defined as the square of this complex amplitude\cite{fe:ls}. For a register with \(M\) qubits, 
the simulator uses a vector space of dimension \(2^{M}\).

\subsection{Quantum Transformations and Logic Gates}
A quantum computation is a sequence of transformations performed on the qubits contained in 
quantum registers\cite{feyn}\cite{ba:be}. A transformation 
takes an input 
quantum state and produces a modified output quantum state. Typically we define transformations 
at the gate level, i.e. transformations which perform logic functions. The simulator performs 
each transformation by multiplying the \(2^{M}\) dimensional vector by a 
\(2^{M} \times 2^{M}\)
transformation matrix. 

The basic gate used in quantum computation is the controlled-not, i.e. exclusive or gate. 
The controlled-not gate is a two bit operation between a control bit and a resultant bit. 
The operation of the gate leaves the control bit unchanged, but conditionally flips the 
resultant bit based on the value of the control bit. 

%

\subsection{The Ion Trap Quantum Computer}
The ion trap quantum computer as proposed by Cirac and Zoller is one of the most promising 
schemes for the experimental realization of a quantum computer \cite{ci:zo}. 
Several experiments have demonstrated simple quantum gates \cite{mo:me} \cite{wi:mm}.
Laser pulses directed at the ions in the trap cause transformations 
to their internal state. A common phonon vibration mode is used to communicate between
the ions in the trap.
A controlled-not gate is a sequence of laser pulses. We use 
the ion trap quantum computer as the model for our quantum simulator.

\subsubsection{Qubits in the Ion Trap Quantum Computer.}
Qubits are represented using the internal energy states of the ions in the trap. 
The ion trap represents a logic zero with the ground state of an ion, and a logic one 
with a higher energy state. The ion trap quantum computer also requires a third state 
which it uses to implement the controlled-not gate. In this paper we use a simplified 
model which, instead of using a third state for each qubit, uses a single third state 
which is shared amongst all the qubits. This simplified model reduces the simulation 
complexity exponentially without an appreciable loss of accuracy\cite{ob2}.


\subsubsection{Transformations in the Ion Trap.}

An operation in the ion trap quantum computer is a sequence of laser pulses. Each laser 
pulse is defined by one of the transformation matrices shown in equation \ref{eq:trans}.
$\theta$ corresponds to the duration of the laser pulse and $\phi$ corresponds to the phase. 
A two bit controlled-not gate is a sequence of five laser pulses, two V 
and three U transformations. A single bit not gate can be implemented with three laser 
pulses and the three bit controlled-controlled-not gate requires seven laser pulses.

\begin{equation}
\label{eq:trans}
U =
\left[
\begin{array}{cccc}
1 & 0 & 0 & 0 \\
0 & \cos\frac{\theta}{2} & -ie^{-i\phi}\sin\frac{\theta}{2} & 0 \\
0 & -ie^{i\phi}\sin\frac{\theta}{2} & \cos\frac{\theta}{2} & 0 \\
0 & 0 & 0 & 1 
\end{array}
\right]
V =
\left[
\begin{array}{cc}
\cos\frac{\theta}{2} & -ie^{-i\phi}\sin\frac{\theta}{2} \\
-ie^{i\phi}\sin\frac{\theta}{2} & \cos\frac{\theta}{2} \\
\end{array}
\right]
\end{equation}

\section{Simulating a Quantum Computer}
Our quantum computer simulator simulates circuits at the gate level. The simulator 
implements one, two and three bit controlled-not gates as well as rotation gates. 
The simulator implements each gate as a sequence of laser pulses, and represents the 
entire vector space throughout the simulation.

\subsection{Operational Errors and Decoherence}
The simulator models inaccuracies by adding a small deviation to the two angles of 
rotation $\theta$ and $\phi$. Each {\em operational} error angle is drawn from a gaussian 
distribution with a parametrized mean ($\mu$) and standard deviation ($\sigma$). 
Errors with non zero $\mu$, called {\em mean} errors, correspond to systematic 
calibration errors, and errors with non zero $\sigma$, referred to as {\em standard deviation} 
errors, correspond to noise in the laser apparatus.

Because the phonon mode is coupled to all the qubits in the computer, it is the 
largest source of decoherence\cite{mo:me}. For this reason we only model the phonon 
decoherence and not the decoherence of the individual qubits.

We model the decoherence of the phonon mode by performing an additional operation after 
each laser pulse. Equation \ref{eq:dec} shows this transformation which has the effect of decaying 
the amplitude of the states in the  phonon state. This decay transformation is based on the 
quantum jump approach\cite{carm}. The decay parameter ({\em dec}) remains 
constant throughout the entire simulation.

\begin{equation}
\label{eq:dec}
\begin{array}{lcr}
	\begin{array}{c}
		|\psi>|0>_{p} \\
		|\psi>|1>_{p} 
	\end{array} &
	\Rightarrow &
	\begin{array}{r}
		|\psi>|0>_{p} \\
		e^{-dec/2}|\psi>|1>_{p} 
	\end{array} 
\end{array}
\end{equation}
This method of modeling decoherence implicitly models spontaneous emission. Because the 
state is never renormalized, the total norm at each step represents the probability that 
the calculation survives up to that point without a spontaneous emission occurring. 
An alternative method for modeling decoherence is to renormalize the state at each step 
and then, based on a probability of emission, cause emissions at different points in the 
calculation. This method has the disadvantage that, because we cause emissions at random 
points in the calculation, we must run multiple simulations each with different initial 
random seeds to average out any bias caused by the random number generator. We have 
shown however that both methods for modeling decoherence give essentially the same 
results\cite{ob2}. 

Because the simulator applies the decoherence transformation once per laser pulse, the 
parameter dec has units of (decoherence/laser pulse). To convert these units to decoherence 
per unit time we must consider the switching time of the laser. 
The $dec$ parameter is simply the switching speed divided by the 
decoherence time. Recent experiments show 
switching speeds of 20kHz for a controlled-not gate, i.e. four $\pi$ pulses, and a decoherence 
rate of a few kHz\cite{mo:me}. This corresponds to a decoherence parameter value between 
$10^{-2}$ and $10^{-3}$.


\subsection{Quantum Circuits}
Much of the current interest in quantum computation is due to the discovery of an 
efficient algorithm by Peter Shor to factor numbers\cite{shor}. By putting the qubit register 
$A$ in the superposition of all values and calculating the function $f(A) = X^{A} mod N$, a 
quantum computer calculates all the values of $f(A)$ simultaneously. Where $N$ is the number 
to be factored, and $X$ is a randomly selected number which is relatively prime to $N$. 
The quantum factoring circuit also contains operations to create the superposition state 
at the beginning of the circuit, and extract the period at the end of the circuit.
The circuit to calculate $f(A)$ can be performed in $O(L^{3})$ time using repeated 
squaring\cite{desp}, i.e. a sequence of multiplications performed modulo $N$. 

Grover's database search algorithm searches for a key, from a set of matching keys, in an 
unsorted database\cite{grov}. 
The keys are defined by a function which can be evaluated in unit time. 
After evaluating this function a {\em diffusion} 
transformation is performed which amplifies the probability in the states with matching 
keys. Grover shows that after performing  $O(\sqrt(N))$ evaluation steps and diffusion 
transformations the probability of measuring a matching key is greater than 1/2. 


\section{Simulation Results}
In this section we study how decoherence and operational errors degrade the fidelity 
of the factoring and database search algorithms. 

The fidelity, as defined by $fidelity = \|<\varphi|\psi>\|^{2}$, measures how close a state 
with error in it is to the correct result. The fidelity is defined as the inner product 
between the simulation with errors ($\psi$) and the correct result ($\varphi$). 

Table \ref{tbl:bench} shows all the benchmarks used in our simulation studies.
To show the complexity of simulating these benchmarks we show the simulation time for each. 
This time assumes a single 300 MHz processor and the simulations include only operational 
error. All the factoring benchmarks were run using a parallel version of the 
simulator\cite{ob3}. We have run simulations on as many as 256 processors, and the 
simulator achieves near linear speedup.

\begin{table}[htbp]
\caption{Benchmarks used in simulation studies}
\label{tbl:bench}
\begin{center}
\begin{tabular}{|c|c|c|c|c|} \hline
\makebox[0.73in]{} & \makebox[0.7in]{{\bf Number of}} & \makebox[0.75in]{{\bf Number of}} & 
\makebox[1.7in]{} &
\makebox[0.72in]{{\bf Simulation}} \\ 
{\bf Benchmark} & {\bf qubits} & {\bf laser pulses} & {\bf Description} & {\bf time (secs)} \\ \hline
& & & Circuit SAT using the Grover &\\ 
grover & 13 & 1,838 & database search algorithm & 10 \\ \hline
& & &  One modulo multiply step &\\ 
mult & 16 & 8,854 & from the factor-15 problem & 282 \\ \hline
& & & Factor-15 problem using &\\ 
factor15 & 18 & 70,793 & 3 qubits for A & 10,465 \\ \hline
& & & Factor-21 problem using &\\ 
factor21 & 24 & 69,884 & 6 qubits for A & 272,276 \\ \hline
& & & Factor-35 problem using &\\ 
factor35 & 27 & 99,387 & 6 qubits for A & 3,083,520 \\ \hline
& & & Factor-57 problem using &\\ 
factor57 & 27 & 97,939 & 6 qubits for A & 3,067,853 \\ \hline
\end{tabular}
\end{center}
\end{table}

Because of round off error, in the factoring algorithm, the choice of the number of bits 
to use in the $A$ register affects the probability seen after performing the 
FFT\cite{shor}. Shor suggest using $2L + 1$ bits for an $L$ bit factorization, 
but for the numbers used in our studies we can use less. For the factor15 problem the 
period is a power of two, i.e. four, and therefore there is no round off 
error. For the numbers 21, 35 and 57, the probability given by the FFT does not 
increase for more than six bits. Also we are mainly concerned with observing the 
fidelity for these circuits, and the fidelity is always calculated before the FFT.

\subsection{Operational Errors}
In this section we consider operational errors without any decoherence. 
We first investigate the significance of errors in the angle $\phi$, 
by varying the amount of error in the angles $\theta$ and $\phi$ separately. 
In all other simulations we vary the angles $\theta$ and $\phi$ together.

To average the random bias out of simulations with non zero $\sigma$ we run multiple 
simulations each with different initial random seeds. To get a single simulation point 
we run at least four simulations, and in many cases we run more to establish upper and lower 
confidence intervals for the average of the ending fidelities\cite{hi:mo}. For simulations 
which include errors in the angle $\theta$ we run enough simulations so that the average of the 
fidelities is within 0.02 of the actual mean of the distribution with an upper and 
lower confidence of 95\%. There is more variability in the fidelity for simulations 
which consider only $\phi$ errors, so for these we obtain the 95\% confidence intervals to 
within 0.03 of the mean.

\subsubsection{Significance of Errors in the Angle $\phi$.}
\begin{figure}[htbp]
\centering \leavevmode \epsfxsize=12.2cm \epsffile{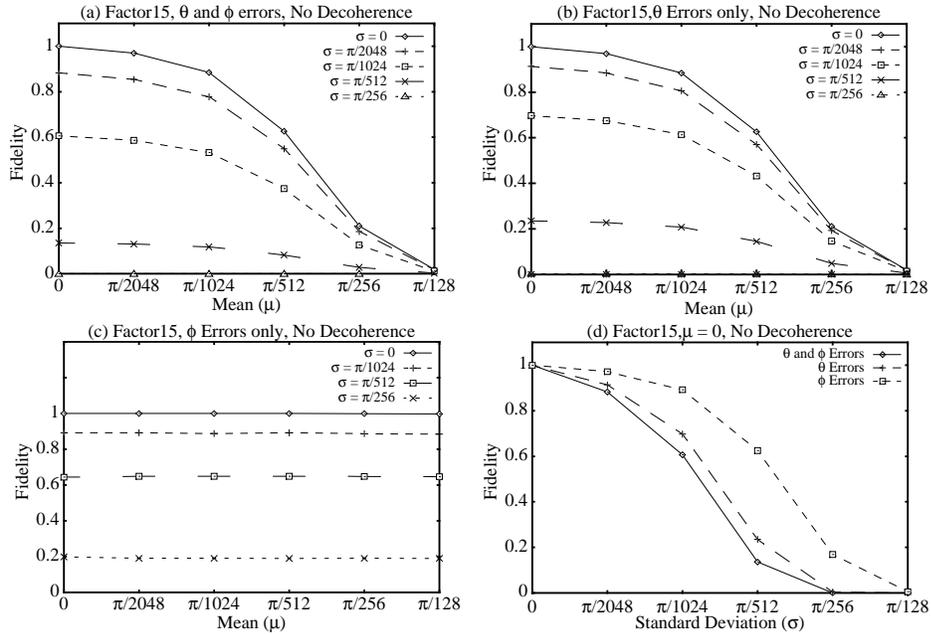}
\caption{Inaccuracies for factor15 (a) $\theta$ and $\phi$ errors (b) $\theta$ errors
(c) $\phi$ errors (d) $\theta$ and $\phi$ errors, $\theta$ errors, $\phi$ errors}
\label{fig:f15op}
\end{figure}

Fig. \ref{fig:f15op} shows how operational errors degrade the fidelity for the factor15 
benchmark. In Fig. \ref{fig:f15op}(a) we vary the mean and standard deviation and 
introduce both $\theta$ and 
$\phi$ errors. In Fig. \ref{fig:f15op}(b) we only introduce $\theta$ errors. 
Comparing the two graphs shows that the combination of $\theta$ and 
$\phi$ errors produces a lower fidelity than $\theta$ errors alone.

Fig. \ref{fig:f15op}(c) 
shows the effect of only $\phi$ errors. As the graph shows noise degrades 
the fidelity, but adding a constant amount of error has no effect. Fixed magnitude $\phi$ 
errors have no effect because the laser transformations are always performed in pairs, 
and an error in the second transformation cancels an error in the first transformation. 
In Fig. \ref{fig:f15op}(d) we compare the effect of $\theta$ and $\phi$ errors alone and their 
combined effect. The highest degradation occurs when considering both $\theta$ and $\phi$ 
errors, and $\theta$ errors produce a more significant effect than $\phi$ errors. 

\subsubsection{Operational Errors for the Grover Benchmark.}
\begin{figure}[htbp]
\centering \leavevmode \epsfxsize=12.2cm \epsffile{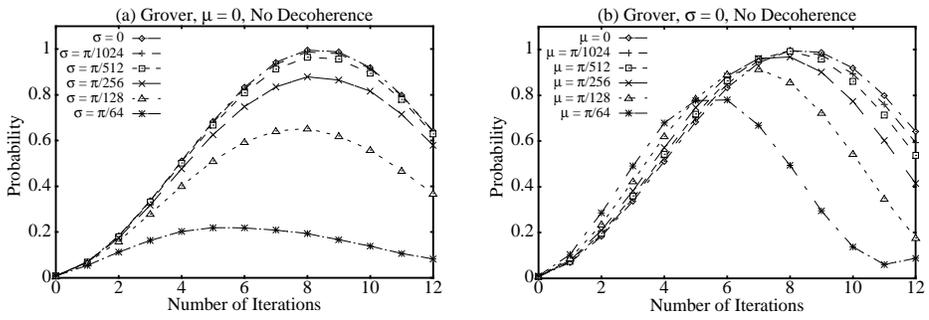}
\caption{Inaccuracies for the grover benchmark. (a) $\sigma$ Errors. (b) $\mu$ errors}
\label{fig:grovop}
\end{figure}

Fig. \ref{fig:grovop}
shows mean and standard deviation operational errors for the grover benchmark. 
The figures show the probability of finding a correct key for twelve iterations of the 
algorithm. When running the database search it is important to stop after the correct 
iteration, because the probability decreases if we run too many iterations\cite{bo:br}. 
For our case the highest probability occurs after the eighth iteration.

Fig. \ref{fig:grovop}(a) 
shows that the peak probability is above 0.5 for $\sigma$ errors as great 
as $\pi/128$, and for an error rate of $\pi/64$ the peak probability is 0.2. 
Also for $\sigma$ errors the peak probability occurs after the same iteration for all 
errors less than $\pi/128$. However $\mu$ errors, as shown in Fig. \ref{fig:grovop}(b), 
shift the
peak so that it occurs at an earlier iteration. The peaks are higher for $\mu$ errors than 
they are for $\sigma$ errors with the same level of error. However, since we can only 
perform a single measurement, the shift in the peak values causes a further reduction in 
the probability if we measure after the eighth iteration.

%

\subsubsection{Fidelity at Intermediate Points for the Factoring Benchmarks.}
\begin{figure}[htbp]
\centering \leavevmode \epsfxsize=12.2cm \epsffile{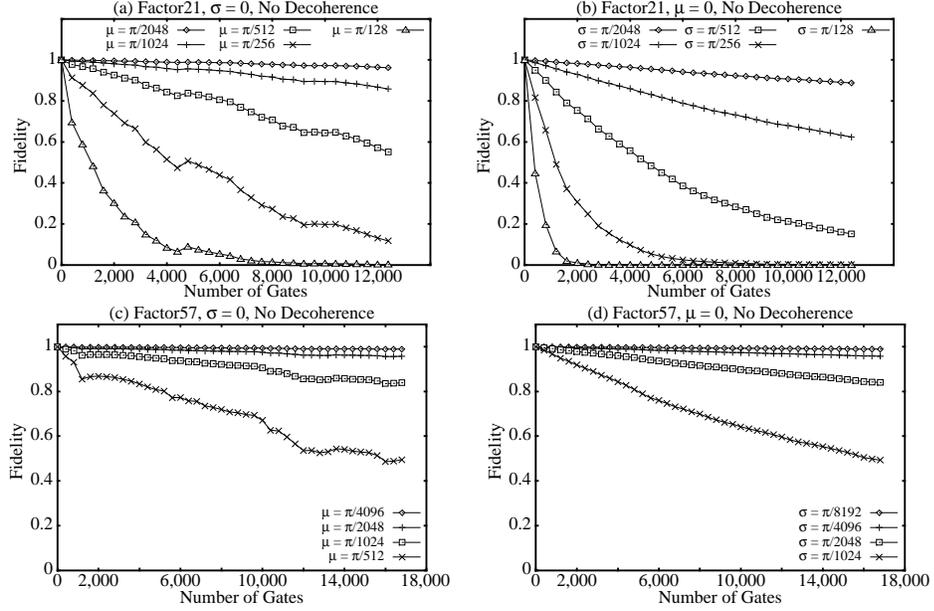}
\caption{Fidelity as a function of the number of gates for factor21 and factor57}
\label{fig:step}
\end{figure}

Fig. \ref{fig:step} shows the fidelity at intermediate points in the calculation for the 
factor21, and factor57 benchmarks.

Standard deviation errors produce a more significant effect than $\mu$ errors of the same 
magnitude. This is due to a cancellation effect for mean errors which is very similar to 
the cancellation effect exhibited by $\phi$ errors\cite{ob2}. This cancellation effect 
occurs because of the nature of reversible computation and causes the fidelity to go 
through periods where it increases. Because we perform operations in pairs, where the 
two operations are inverses of each other, an error in one operation may reverse the error 
from an earlier operation. Standard deviation errors also exhibit this effect 
but it is reduced because we average the results from multiple simulations.

\subsection{Decoherence Errors}
\begin{figure}[htbp]
\centering \leavevmode \epsfxsize=12.2cm \epsffile{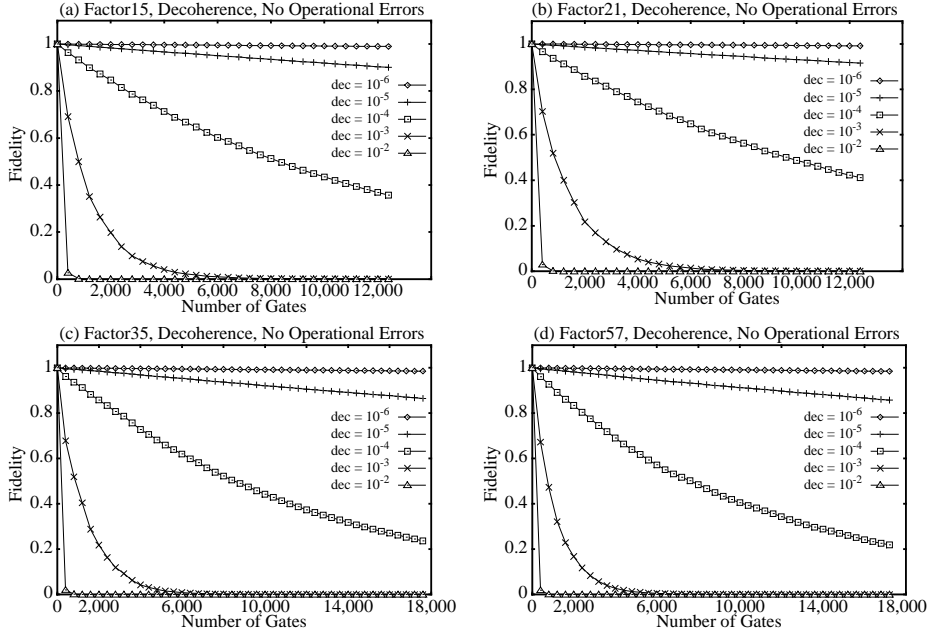}
\caption{Fidelity as a function of the number of gates for decoherence in the factor15, factor21, factor35 and factor57 benchmarks}
\label{fig:dec}
\end{figure}

Fig. \ref{fig:dec}
shows how the fidelity of a computation decreases over time in the presence of 
decoherence. Fig. \ref{fig:dec} shows the fidelity at intermediate points in the calculation for 
the four factoring benchmarks. For small amounts of decoherence, i.e. $10^{-6}$ or less, 
the fidelity is not adversely affected. A decoherence rate of $10^{-4}$ results in a steady 
decrease in the fidelity over the course of the computation, and for rates even higher the 
fidelity drops off very quickly. As the figure shows, decoherence has a similar effect on
all of the benchmarks.


\subsection{The Correlation Between Decoherence and Operational Errors}
Both decoherence and operational error cause a degradation of the fidelity in a quantum 
computation. Decoherence degrades the fidelity through the decay of the phonon state, 
and operational errors result in the accumulation of amplitude in unwanted states. 
The combined effect of these two factors is a degradation which is worse than either 
factor considered alone. We can represent the combined effect as:
$F_{dec,op} = F_{dec} \bullet F_{op} + \Omega(F_{dec},F_{op})$
Where $F_{dec}$ and $F_{op}$ 
are the fidelities of simulations for decoherence and operational error considered 
separately, and  $\Omega(F_{dec},F_{op})$ is the correlation between the two types of 
error. As Table \ref{tbl:corr} shows the correlation is very low. We calculated the correlation 
by running simulations which considered decoherence and operational error together. 
For all the benchmarks the maximum correlation is at most $1.14\times10^{-2}$. This result means 
that we can simulate decoherence and operational errors separately, and combine the 
results to obtain their collective effect on a calculation. 

\begin{table}[htbp]
\caption{Correlation ($\Omega$) between decoherence and operational errors}
\label{tbl:corr}
\begin{center}
\begin{tabular}{|c|c|c|} \hline
{\bf Benchmark and Simulation Model} & {\bf Maximum $\Omega$} & {\bf Average $\Omega$} \\ \hline
\rule{0pt}{10pt}mult, $\mu$=0, $\sigma$ = $\pi/1024$ - $\pi/64$ & $5.76\times10^{-5}$ & $3.63\times10^{-6}$  \\ \hline
\rule{0pt}{10pt}mult, $\sigma$ = 0, $\mu$ = $\pi/1024$ - $\pi/64$ & $9.26\times10^{-3}$ & $5.51\times10^{-4}$ \\ \hline
\rule{0pt}{10pt}factor15, $\sigma$ = $\pi/1024$, $\mu$ = 0 & $4.15\times10^{-3}$ & $3.96\times10^{-4}$ \\ \hline
\rule{0pt}{10pt}factor15, $\sigma$ = 0, $\mu$ = $\pi/1024$ & $1.14\times10^{-2}$ & $8.76\times10^{-4}$ \\ \hline
\rule{0pt}{10pt}grover, $\mu$=0, $\sigma$ = $\pi/1024$ - $\pi/128$  & $1.78\times10^{-3}$ & $1.02\times10^{-4}$ \\ \hline
\rule{0pt}{10pt}grover, $\sigma$=0, $\mu$ = $\pi/1024$ - $\pi/128$ & $2.67\times10^{-3}$ & $2.36\times10^{-4}$ \\ \hline
\end{tabular}
\end{center}
\end{table}

The complexity of simulating decoherence alone is much lower than the complexity of simulating
operational errors. This is because the decay transformation does not have any off diagonal 
terms, and therefore it does not introduce any new error states. The simulator only needs to 
represent enough states to represent the superposition state. Instead of using the index of 
a state to represent its bit value, the simulator now keeps an extra field for each state 
which holds the current value of the bit string for that state. 

Using this new method for modeling decoherence, the simulator only needs to allocate $O(2^{L})$ 
states to simulate the factorization of an $L$ bit number. To represent all the qubits for 
this problem, the simulator would need to allocate $O(2^{4L})$ states. This reduces the 
memory requirements and simulation complexity by a factor of $O(2^{3L})$. For example the 
simulation of the factor15 circuit requires only 1/4096 the amount of time as before.

\subsection{The Error Rate per Gate}
As our simulation results show, a modest amount of error destroys even a relatively small 
calculation. If quantum computers are to be useful, we must be able to perform calculations 
which are even larger than the ones considered here. These larger calculations will therefore 
require the use of quantum error correcting codes\cite{stea}. Several recent studies have 
shown that, by using fault tolerant techniques, if the error of an individual gate is low 
enough we can perform a useful quantum calculation of indefinite length\cite{pres}\cite{kn:lz}.
This {\em accuracy threshold} is expressed in terms of the probability of 
error per gate. We can use the results of our simulation studies to show how this error 
probability relates to decoherence and inaccuracies. 

Table \ref{tbl:rate} shows the error rate per gate considering decoherence and operational errors 
for the factor57 benchmark. Error rates for the other factoring benchmarks as well as 
error rates for the combination of decoherence and operational errors are given in \cite{ob4}.
 The error rate is calculated as {\em (1 - Fidelity)/Number of Gates}. To get the error 
rate for a particular amount of error, we calculate the error rate after every tenth 
gate in a computation and take the average. A gate is either a one, two or three bit 
controlled-not gate or a single bit rotation. It takes five laser pulses on average to 
implement each gate.

\begin{table}[htbp]
\caption{Average error rate per gate for the factor57 benchmark}
\label{tbl:rate}
\begin{center}
\begin{tabular}{|c|c||c|c||c|c|} \hline
	\multicolumn{2}{|c||}{{\bf Decoherence}}&\multicolumn{2}{|c||}{{\bf Operational Errors, $\mu=0$}}&\multicolumn{2}{|c||}{{\bf Operational Errors, $\sigma=0$}} \\ \hline
	dec & Error Rate & $\sigma$ & Error Rate & $\mu$ & Error Rate \\ \hline
\rule{0pt}{10pt}	$10^{-7}$ & $9.1\times10^{-8}$ & $\pi/8192$ & $6.6\times10^{-7}$ & $\pi/4096$ & $7.9\times10^{-7}$ \\ \hline
\rule{0pt}{10pt}	$10^{-6}$ & $9.1\times10^{-7}$ & $\pi/4096$ & $2.6\times10^{-6}$ & $\pi/2048$ & $3.1\times10^{-6}$ \\ \hline
\rule{0pt}{10pt}	$10^{-5}$ & $8.8\times10^{-6}$ & $\pi/2048$ & $1.0\times10^{-5}$ & $\pi/1024$ & $1.2\times10^{-5}$ \\ \hline
\rule{0pt}{10pt}	$10^{-4}$ & $6.5\times10^{-5}$ & $\pi/1024$ & $3.6\times10^{-5}$ & $\pi/512$ & $4.4\times10^{-5}$ \\ \hline
\end{tabular}
\end{center}
\end{table}


To perform a computation of arbitrary length the error rate must be about $10^{-5}$ for one 
and two bit gates and $10^{-3}$ for three bit gates[Pres96]. An error rate of $10^{-5}$ 
corresponds to operational errors with $\sigma=\pi/2048$ and decoherence of $10^{-5}$. 
Table \ref{tbl:rate} shows that we can tolerate an even higher level of constant magnitude errors. 
The error rate is $10^{-5}$ for operational errors with $\mu=\pi/1024$.

To perform a quantum factorization which is more efficient than a classical one the error 
threshold is even tighter, roughly $10^{-6}$. This corresponds to operational errors 
with $\sigma<\pi/4096$ and a decoherence rate of $10^{-6}$.

Using the error rate of our factoring circuits to predict the error rate for error correction 
circuits assumes that these two types of circuits behave in a similar manner. These circuits 
are similar because they are both built from the same types of elementary gates, i.e. 
controlled-not gates. Also the error rate, for a given amount of error, is very similar 
for all the factoring circuits that we considered. Lastly effects that we have seen such 
as error cancellation are a by-product of the fact that quantum circuits and gates are 
implemented in a reversible fashion. Because of the nature quantum mechanics quantum 
circuits will always be implemented in this way.

%
%

\section{Conclusion}
Quantum computation is a new type of computation which can achieve exponential parallelism. 
The feasibility of a quantum computer is threatened by two types of errors, decoherence 
and inaccuracies. In this paper we performed simulations of a quantum computer running 
Shor's factoring algorithm and Grover's database search algorithm to access the feasibility 
of implementing a quantum computer. 

Our simulations show that random inaccuracies (noise) are more significant than fixed magnitude 
inaccuracies for the ion trap quantum computer. Also errors in the duration of the laser 
pulse are more significant than errors in the phase of the laser. For the problems considered 
in this paper we show that noise or constant magnitude inaccuracies of magnitude $\pi/512$ 
or greater cause a significant impact on the fidelity of the calculation.

Our simulations also show that a quantum computation can tolerate a decoherence rate as high as 
$10^{-5}$. For the ion trap computer this corresponds roughly to a decoherence lifetime of 
50 milliseconds and a switching speed of 500 nanoseconds. We also show that inaccuracies 
and decoherence are uncorrelated and can be simulated separately.

Our simulations relate the physical quantities of inaccuracies and decoherence to the 
probability of error per gate. An error rate per gate on the order of $10^{-6}$ corresponds 
to inaccuracies of less than $\pi/4096$ per laser operation and a decoherence rate of 
$10^{-6}$. A quantum computer with this error rate, if it uses quantum error correcting 
codes, could factor a number more efficiently than a classical computer. This assumes that 
the quantum circuits used to implement factoring with error correction codes behave in the 
same manner as the factoring circuits used in this paper.

\subsection*{Acknowledgments}
The authors are members of the Quantum Information and Computation (QUIC) consortium. We 
wish to thank our QUIC colleagues: Jeff Kimble, John Preskill, Hideo Mabuchi and Dave Vernooy. 
This work is supported in part by ARPA under contract number DAAH04-96-1-0386.

\end{document}